\documentclass[preprintnumbers]{revtex4}
\usepackage{amssymb}
\usepackage{graphicx,amsmath}
\usepackage{bm}
\usepackage{times}

\begin{document}

\title{Quantum behaviour of a flux qubit coupled to a resonator}

\begin{abstract}
We present a detailed theoretical analysis for a system of a superconducting
flux qubit coupled to a transmission line resonator. The master equation,
accounting incoherent processes for a weakly populated resonator, is
analytically solved. An electromagnetic wave transmission coefficient
through the system, which provides a tool for probing dressed states of the
qubit, is derived. We also consider a general case for the resonator with
more than one photon population and compare the results with an experiment
on the qubit-resonator system in the intermediate coupling regime, when the
coupling energy is comparable with the qubit relaxation rate.

Keywords: superconducting qubit, transmission line, resonator.
\end{abstract}

\date{\today }
\author{A. N. Omelyanchouk}
\affiliation{B. Verkin Institute for Low Temperature Physics and Engineering, 47 Lenin
Ave., 61103 Kharkov, Ukraine}
\author{S. N. Shevchenko}
\email{sshevchenko@ilt.kharkov.ua}
\affiliation{B. Verkin Institute for Low Temperature Physics and Engineering, 47 Lenin
Ave., 61103 Kharkov, Ukraine}
\affiliation{Institute of Photonic Technology, P.O. Box 100239, D-07702 Jena, Germany}
\author{Ya. S. Greenberg}
\affiliation{Novosibirsk State Technical University, 20 Karl Marx Ave., 630092 Russia}
\affiliation{Institute of Photonic Technology, P.O. Box 100239, D-07702 Jena, Germany}
\author{O.~Astafiev}
\affiliation{NEC Nano Electronics Research Laboratories, Tsukuba, Ibaraki, 305-8501, Japan}
\author{E.~Il'ichev}
\affiliation{Institute of Photonic Technology, P.O. Box 100239, D-07702 Jena, Germany}
\pacs{85.25.Am, 85.25.Cp, 84.37.+q}
\maketitle

\section{Introduction}

Modern state of the art fabrication using nanotechnology brings together
quantum optics and mesoscopic solid state physics. Different types of
Josephson-junction quantum bits (qubits) -- macroscopic quantum objects --
are now intensively studied and their quantum behavior have been
experimentally demonstrated (for review see e.g.\cite{Ilich, You05,
Wen-Shum07, Zagoskin}). A series of quantum phenomena such as, for example,
entanglement \cite{Pash, Izmalk}, Rabi oscillations \cite{Vion05, Chi1,
Chi2, Ilich1, Johan, Omel}, spin-echo and Ramsey fringes \cite{Nakam, Vion},
Landau-Zener-St\"{u}ckelberg interferometry \cite{Izmalk1, Oliver,
Sillanpaa, Shevchenko10, Sun10}, have been recently demonstrated. Now, great
interest is attracted to physics of artificial atoms (built on the basis of
qubits) in a confined fields of electromagnetic resonators, which is known
as circuit quantum electrodynamics (CQED) \cite{Schoel}. In pioneering CQED
experiments, the artificial atom was electrostatically coupled to a
high-quality transmission line resonator. The large electrical dipole moment
of the qubit and high energy density of the resonator allow this system to
reach the strong coupling limit. %\textit{\ }Coupling the superconducting
%qubits to the high-quality resonators can be realized with the \textit{%
%distributed elements such as} the transmission-line resonators.
This regime was studied theoretically \cite{Blais04, Liu06, Chen, Bour,
Ashhab09} and experimentally for the charge qubit coupled capacitively to
the resonator \cite{Wall, Schuster, Schuster1}. Later inductive coupling for
the flux qubit was proposed \cite{Lindstrom07} and realized experimentally
\cite{Abdumalikov08,Oelsner09}.

In this paper, we analyze the system of the superconducting flux qubit
coupled to the transmission line resonator. Our aim is first to present the
detailed theory of the qubit's states, dressed by the interaction with the
quantum resonator, and their influence on the observable transmission.
Second, we describe the regime of intermediate coupling studied recently
experimentally by Oelsner et al. \cite{Oelsner09}. Accordingly, the paper is
organized as follows. In the next Section the model of the system is
described. In Sec. III, we calculate energy levels of the system. Allowed
transitions between the levels can be measured by spectroscopy applying
external driving fields. Different representations of the system Hamiltonian
are discussed in Sec. IV, particularly, the rotating-wave approximation
(RWA), convenient for finding the stationary solutions. The analytical
solution for the master equation is presented in Sec. V for the case of a
weak and numeric calculations for strong driving regimes. In Appendix we
present details of the theory for the transmission through the resonator.

\section{Description of the system}

\begin{figure}[h]
\includegraphics[width=8cm]{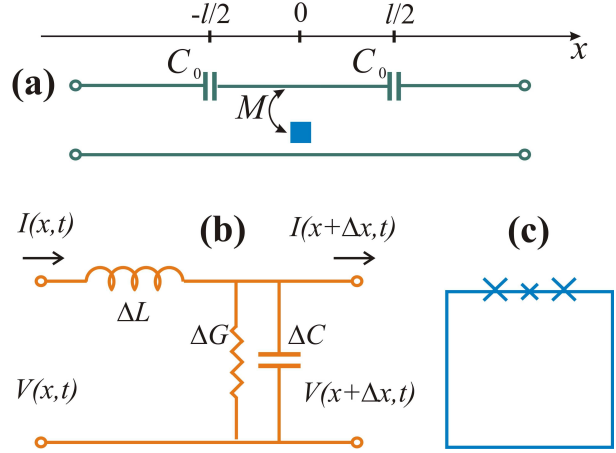}
\caption{(a) Schematics of the qubit (denoted by a blue box) coupled to the
transmission line resonator via inductance $M$. (b) Equivalent circuit for
the description of the infinitesimal piece of length $\Delta x$ of the
transmission line. (c) Flux qubit with $3$ Josephson junctions.}
\label{Fig:scheme}
\end{figure}

We consider the flux qubit coupled inductively to a coplanar waveguide
resonator, see Fig.~\ref{Fig:scheme}. The flux qubit is a superconducting
loop with three Josephson junctions \cite{Orlando99}. Two qubit states are
naturally described in a flux basis. The two flux states ($|\uparrow \rangle
$, $|\downarrow \rangle $) are differed by directions of the circulating
current (clockwise and counterclockwise) in the loop. The qubit current
interacts with the field of the resonator. The coplanar waveguide resonator
is defined by two gaps in the transmission line, which form capacitances $%
C_{0}$ at $x=\pm l/2$. The qubit is situated close to the center of the
resonator ($x=0$), where the current of the resonator fundamental mode is
maximal. Note that the qubit dimensions are significantly smaller than the
resonator wavelength, therefore we consider it as a point-like object.

The total Hamiltonian of the driven system
\begin{equation}
H=H_{\mathrm{qb-r}}+H_{\mathrm{\mu w}}  \label{Htot}
\end{equation}
is a sum of the driving field Hamiltonian $H_{\mu \mathrm{w}}$ and the
qubit-resonator Hamiltonian
\begin{equation}
H_{\mathrm{qb-r}}=H_{\mathrm{qb}}+H_{\mathrm{r}}+H_{\mathrm{int}},
\label{H_qb-r}
\end{equation}%
which consists of the bare qubit $H_{\mathrm{qb}}$, resonator $H_{\mathrm{r}%
} $ and the interaction term $H_{\mathrm{int}}$. The flux qubit Hamiltonian
in the flux basis has the form \cite{Orlando99}

\begin{equation}
H_{\mathrm{qb}}=-\frac{\Delta }{2}\sigma _{x}-\frac{\varepsilon }{2}\sigma
_{z},  \label{Hqb}
\end{equation}%
where $\Delta $ is the tunnelling amplitude, the energy bias $\varepsilon
=2I_{\mathrm{p}}(\Phi -\Phi _{0}/2)$ is defined by the magnetic flux $\Phi $%
, $I_{\mathrm{p}}$ is the persistent current, $\sigma _{x,z}$ are the Pauli
matrices ($\sigma _{z}\left\vert \downarrow \right\rangle =-\left\vert
\downarrow \right\rangle $); the current operator is $\hat{I}_{\mathrm{qb}%
}=-I_{\mathrm{p}}\sigma _{z}$.

The qubit is coupled to the transmission line resonator. A detailed
resonator description is presented in Appendix (see also Refs. \cite%
{Oelsner09,Astafiev10,Zhou08}). The single-mode resonator is described by
the following Hamiltonian

\begin{equation}
H_{\mathrm{r}}=\hbar \omega _{\mathrm{r}}\left( a^{\dag }a+\frac{1}{2}%
\right) ,  \label{Hosc}
\end{equation}%
where $a$ and $a^{\dag }$ are the annihilation and creation operators, which
act at the number (Fock) states according to $a\left\vert n\right\rangle =%
\sqrt{n}\left\vert n-1\right\rangle $ and $a^{\dag }\left\vert
n-1\right\rangle =\sqrt{n}\left\vert n\right\rangle $.

The term, describing the interaction between the resonator and the flux
qubit, is

\begin{equation}
H_{\mathrm{int}}=M\hat{I}(0)\hat{I}_{\mathrm{qb}}=-\hbar \mathrm{g}(a^{\dag
}+a)\sigma _{z},  \label{Hint}
\end{equation}%
\begin{equation}
\hbar \mathrm{g}=MI_{\mathrm{r}0}I_{\mathrm{p}},  \label{g}
\end{equation}%
where $M$ is the mutual loop-resonator inductance, $\hat{I}(0)= I_{\mathrm{r0%
}}(a+a^\dag)$ is the transmission line current operator (Eq.~(\ref{I_TL_oper}%
)), at the qubit's position, $x=0$.

The transmission line resonator is driven by the external probing voltage field at the frequency $\omega_\mathrm{d}$ close to the resonator
characteristic frequency $\omega _{\mathrm{r}}$, as described in the
Appendix. The qubit in turn is driven by the resonator photon field with the
amplitude $\xi $ and the frequency $\omega _{\mathrm{d}}$. The Hamiltonian
of this field, described by photon exchange between the resonator and the
driving field, can be written as
\begin{equation}
H_{\mathrm{\mu w}}=\xi \left(iae^{i\omega _{\mathrm{d}}t} - ia^{\dag
}e^{-i\omega _{\mathrm{d}}t}\right)  \label{Hmw}
\end{equation}
(the derivation is presented in Sec. IV)

\section{Energy levels and the spectroscopy of dressed states}

\begin{figure}[h]
\includegraphics[width=8cm]{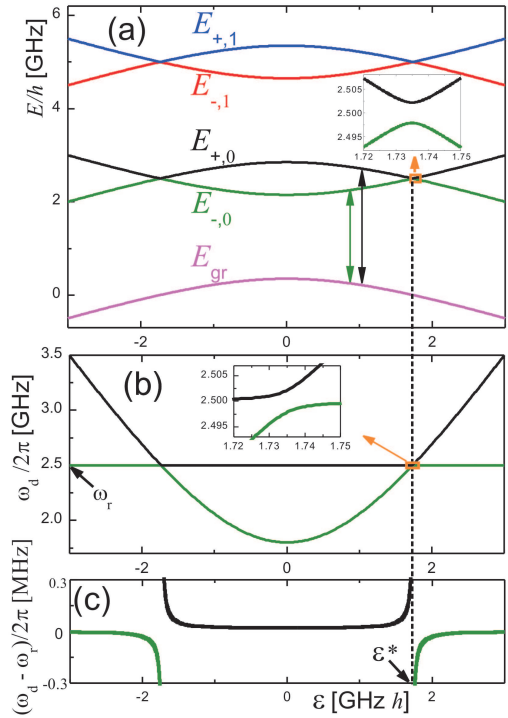}
\caption{Energy diagrams calculated for the following qubit parameters $%
\Delta /h=1.8$ GHz, $\mathrm{g}/2\protect\pi =3$ MHz, $\protect\omega _{%
\mathrm{r}}/2\protect\pi =2.5$ GHz. (a) Energy levels versus energy bias $%
\protect\varepsilon $. Avoided level crossing is shown as a close-up in the
inset. (b) Contour lines of the energy difference versus bias $\protect%
\varepsilon $ and the driving frequency $\protect\omega _{\mathrm{d}}$.
The green (lower) line is for $\hbar \protect\omega _{\mathrm{d}}=E_{-,0}-E_{%
\mathrm{gr}}$ and the black (upper) line is for $\hbar \protect\omega _{%
\mathrm{d}}=E_{+,0}-E_{\mathrm{gr}}$. (c) The same plot as in (b) but in a
narrow vicinity to the resonator fundamental frequency $\protect\omega _{%
\mathrm{r}}$.}
\label{Fig:levels}
\end{figure}

The qubit-resonator Hamiltonian in the qubit eigenbasis (see, e.g., Ref.
\cite{Greenberg07}) can be written as%
\begin{equation}
H_{\mathrm{qb-r}}^{\prime }=H_{0}^{\prime }+H_{\mathrm{int}}^{\prime },
\label{Hqbr}
\end{equation}%
where
\begin{equation}
H_{0}^{\prime }=\frac{\hbar \omega _{\mathrm{qb}}}{2}\sigma _{z}+\hbar
\omega _{\mathrm{r}}\left( a^{\dag }a+\frac{1}{2}\right) ,  \label{H0}
\end{equation}%
\begin{equation}
H_{\mathrm{int}}^{\prime }=-\hbar \mathrm{g}(a^{\dag }+a)\left( \frac{%
\varepsilon }{\hbar \omega _{\mathrm{qb}}}\sigma _{z}-\frac{\Delta }{\hbar
\omega _{\mathrm{qb}}}\sigma _{x}\right) ,  \label{Hint2}
\end{equation}%
with the bare qubit energy splitting
\begin{equation}
\hbar \omega _{\mathrm{qb}}=\sqrt{\Delta ^{2}+\varepsilon ^{2}}.  \label{wqb}
\end{equation}%
The bare system eigenstates are $\left\vert e/g,n\right\rangle =\left\vert
e/g\right\rangle \otimes \left\vert n\right\rangle $ and eigenvalues%
\begin{equation}
E_{e/g,n}=\pm \frac{\hbar \omega _{\mathrm{qb}}}{2}+\hbar \omega _{\mathrm{r}%
}\left( n+\frac{1}{2}\right) .  \label{Epm}
\end{equation}%
The states $|e,n\rangle $ and $|g,n+1\rangle $ are degenerated at $\omega _{%
\mathrm{r}}=\omega _{\mathrm{qb}}$ and the degeneracy is lifted by the
qubit-resonator interaction. The transition matrix element due to the
interaction is
\begin{equation}
\left\langle g,n+1\left\vert H_{\mathrm{int}}^{\prime }\right\vert
e,n\right\rangle =\left\langle e,n\left\vert H_{\mathrm{int}}^{\prime
}\right\vert g,n+1\right\rangle =\hbar \mathrm{g}_{\varepsilon }\sqrt{n+1},
\label{off-diagonal}
\end{equation}%
where the qubit-resonator interaction energy is
\begin{equation}
\hbar \mathrm{g}_{\varepsilon }=\hbar \mathrm{g}\frac{\Delta }{\hbar \omega
_{\mathrm{qb}}}.  \label{g_eps}
\end{equation}%
Note that the coupling strength is scaled as $\Delta /\hbar \omega _{\mathrm{%
qb}}$ \cite{Blais04}. The eigenvectors $|+\rangle $ and $|-\rangle $ of the
total Hamiltonian $H_{\mathrm{qb-r}}^{\prime }$ are obtained from the
non-interacting qubit-resonator basis by the following transformation
\begin{equation}
\left(
\begin{array}{c}
\left\vert -,n\right\rangle  \\
\left\vert +,n\right\rangle
\end{array}%
\right) =\left(
\begin{array}{cc}
\sin \eta  & \cos \eta  \\
-\cos \eta  & \sin \eta
\end{array}%
\right) \left(
\begin{array}{c}
\left\vert g,n+1\right\rangle  \\
\left\vert e,n\right\rangle
\end{array}%
\right) ,  \label{new_basis}
\end{equation}%
where
\begin{equation}
\tan 2\eta =\frac{2\mathrm{g}_{\varepsilon }\sqrt{n+1}}{\delta },
\label{theta}
\end{equation}%
\begin{equation}
E_{\pm ,n}=\hbar \omega _{\mathrm{r}}\left( n+1\right) \pm \frac{\hbar
\Omega _{n}}{2},  \label{dressed}
\end{equation}%
\begin{equation}
\Omega _{n}=\sqrt{4\mathrm{g}_{\varepsilon }^{2}\left( n+1\right) +\delta
^{2}},  \label{Omega}
\end{equation}%
\begin{equation}
\delta =\omega _{\mathrm{qb}}-\omega _{\mathrm{r}}<0.  \label{delta}
\end{equation}%
The energy of the ground state, $\left\vert g,0\right\rangle $, is given by%
\begin{equation}
E_{\mathrm{gr}}\equiv E_{g,0}=-\frac{\hbar \delta }{2}.  \label{Eg0}
\end{equation}%
Here $\Omega _{n}$ defines the energy difference $E_{+,n}-E_{-,n}=\hbar
\Omega _{n}$. In particular, the energy anticrossing takes place at $\delta
=0$, that is at $\hbar \omega _{\mathrm{qb}}(\varepsilon ^{\ast })=\hbar
\omega _{\mathrm{r}}$, and it is given by
\begin{equation}
\Omega _{n}^{\min }=\Omega _{n}(\varepsilon ^{\ast })=2\mathrm{g}%
_{\varepsilon ^{\ast }}\sqrt{n+1}=2\mathrm{g}\frac{\Delta }{\hbar \omega _{%
\mathrm{r}}}\sqrt{n+1}.  \label{Omega_min}
\end{equation}%
For example, in the inset of Fig.~\ref{Fig:levels}(a), the energy
anticrossing is shown for $n=0$.

%\begin{figure}[h]
%\includegraphics[width=8cm]{Fig1.eps}
%\caption{} \label{Fig:scheme}
%\end{figure}

If the resonator is driven by a weak external field, so that it is weakly
populated, one can limit the consideration by a few Fock states, neglecting
unpopulated levels; the energy levels are plotted with Eqs.~(\ref{dressed})
and (\ref{Eg0}) in Fig.~\ref{Fig:levels}(a). With the weak driving field the
spectroscopy of the "dressed" energy levels can be done: the transmission is
resonantly increased when the driving photon energy $\hbar \omega _{\mathrm{d%
}}$ matches the system energy difference of Eqs.~(\ref{dressed}) and (\ref%
{Eg0}), shown by two arrowed lines in Fig.~\ref{Fig:levels}(a) for two
possible transitions. One can plot then the respective energy contour lines
to describe experimental results (see Figs.~\ref{Fig:levels}(b, c)), which
relates to the experimental data presented in Figs. 2 and 3 of Ref. \cite%
{Oelsner09}. With increasing driving amplitude, higher order processes such
as multi-photon transitions may become possible \cite{Abdumalikov08,Fink08}.

\section{Hamiltonian of the system}

\textbf{Jaynes-Cummings Hamiltonian.} Let us rewrite the interaction
Hamiltonian, Eq.~(\ref{Hint2}), by introducing the qubit lowering and
raising operators
\begin{equation}
\sigma^\pm =\frac{1}{2}(\sigma _{x}\pm i\sigma _{y}),  \label{sigmas}
\end{equation}%
so that $\sigma ^{+}\left\vert g\right\rangle =\left\vert e\right\rangle $, $%
\sigma ^{+ }\left\vert e\right\rangle =0$, etc.; then we have%
\begin{equation}
H_{\mathrm{int}}^{\prime }=\hbar \mathrm{g}_{\varepsilon }(a^{\dag }\sigma^-
+a\sigma ^{+})+\hbar \mathrm{g}_{\varepsilon }(a\sigma^- +a^{\dag }\sigma
^{+ })-\hbar \mathrm{g}\frac{\varepsilon }{\hbar \omega _{\mathrm{qb}}}%
(a^{\dag }+a)\sigma _{z}.  \label{Hint3}
\end{equation}%
In vicinity of degeneracy of the states $|e,n\rangle $ and $|g,n+1\rangle$,
the second and the third term in Eq.~(\ref{Hint3}) can be neglected as they
correspond to the processes, which require large extra energy. The first
term together with $H_{0}^{\prime }$ from Eq.~(\ref{H0}) give the
Jaynes-Cummings Hamiltonian%
\begin{equation}
H_{\mathrm{JC}}=\frac{\hbar \omega _{\mathrm{qb}}}{2}\sigma _{z}+\hbar
\omega _{\mathrm{r}}\left( a^{\dag }a+\frac{1}{2}\right) +\hbar \mathrm{g}%
_{\varepsilon }(a^{\dag }\sigma^- +a\sigma ^{+}).  \label{H_JC}
\end{equation}

\textbf{Interaction representation. }We consider $H_{\mathrm{int}}^{\prime }$
in the interaction representation. For this, we note the following relations
(see e.g. \cite{Schleich})%
\begin{equation}
e^{ia^{\dag }a\omega t}ae^{-ia^{\dag }a\omega t}=ae^{-i\omega t},
\label{relation4a}
\end{equation}%
\begin{equation}
e^{i\frac{\omega }{2}t\sigma _{z}}\sigma ^{-}e^{-i\frac{\omega }{2}t\sigma
_{z}}=\sigma ^{-}e^{-i\omega t}.  \label{relation4sigma}
\end{equation}%
Then we obtain%
\begin{equation}
H_{\mathrm{int}}^{I}=e^{\frac{i}{\hbar }H_{0}^{\prime }t}H_{\mathrm{int}%
}^{\prime }e^{-\frac{i}{\hbar }H_{0}^{\prime }t}=\hbar \mathrm{g}%
_{\varepsilon }\left( a\sigma ^{+}e^{i(\omega _{\mathrm{qb}}-\omega _{%
\mathrm{r}})t}+h.c.\right)  \notag
\end{equation}%
\begin{equation}
+\hbar \mathrm{g}_{\varepsilon }\left( a\sigma ^{-}e^{-i(\omega _{\mathrm{qb}%
}+\omega _{\mathrm{r}})t}+h.c.\right) -\hbar \mathrm{g}\frac{\varepsilon }{%
\hbar \omega _{\mathrm{qb}}}\left( ae^{-i\omega _{\mathrm{r}}t}+h.c.\right) .
\label{H_I_inter}
\end{equation}%
In the RWA, when $\omega _{\mathrm{qb}}-\omega _{\mathrm{r}}\ll \omega _{%
\mathrm{qb}}$, the first term is slowly rotating, while the second and third
terms are fast rotating ones. This justifies neglecting these terms.

\textbf{Driving Hamiltoinan. }We will consider scattering of the right propagating
wave $V_{1}^{r}(e^{-ik(x+l/2)+i\omega _{\mathrm{d}}t}+c.c.)$ on the resonator
(see Eq. (\ref{Volt1})), where $V_1^r$ is chosen to be a real amplitude. This wave drives the resonator, which in turn
generates the scattered waves. Using semiclassical approach, we first
consider the driving dynamics under the classical field and then calculate
the field generated by the resonator. The calculated first order scattering
gives an exact solution because the two scattered waves, propagating in
different directions cancel out the second order driving effect (see Eq. ({%
\ref{Hmus2})). The dipole-like interaction Hamiltonian can be presented as a
product of the incident wave voltage field and charges generated by the
resonator field on the coupling capacitances $C_{0}V_{\mathrm{r0}%
}(ia-ia^{\dag })\sin {(\pm k_{r}l/2)}$ (see Eq. (\ref{V_TL_oper2}))
\begin{equation}
H_{\mathrm{\mu w}}=\xi \left( e^{i\omega _{\mathrm{d}}t}+e^{-i\omega _{%
\mathrm{d}}t}\right) \left( ia-ia^{\dag }\right) ,  \label{Hmus}
\end{equation}%
where
\begin{equation}
\xi=C_{0}V_{1}^{r}V_{\mathrm{r0}}\sin {(k_{r}l/2)}
(-1+e^{-ik_{r}l}) =-2C_{0}V_{1}^{r}V_{\mathrm{r}0}.  \label{xi}
\end{equation}%
And omitting fast rotating terms in RWA, we arrive to Eq. (\ref{Hmw}). In
these equations we assume $\theta _{1}=\omega C_{0}Z_{1}\ll 1$, which is
valid for high quality resonators.
}

\textbf{Rotating-wave approximation. }We consider the Hamiltonian of the
\emph{driven} system in the RWA%
\begin{equation}
H_{\mathrm{RWA}}=U\left( H_{\mathrm{qb-r}}^{\prime }+H_{\mathrm{\mu w}%
}\right) U^{\dag }+i\hbar \dot{U}U^{\dag }.  \label{H_RWA_1}
\end{equation}%
For this we choose the transformation%
\begin{equation}
U=\exp \left[ i\omega _{\mathrm{d}}t\left( a^{\dag }a+\sigma _{z}/2\right) %
\right]   \label{U_RWA}
\end{equation}%
and obtain%
\begin{equation}
H_{\mathrm{RWA}}=\hbar \frac{\delta \omega _{\mathrm{qb}}}{2}\sigma
_{z}+\hbar \delta \omega _{\mathrm{r}}a^{\dag }a+\hbar \mathrm{g}%
_{\varepsilon }\left( a\sigma ^{+}+a^{\dag }\sigma ^{-}\right) +\xi
(ia-ia^{\dag }),  \label{H_RWA_2}
\end{equation}%
\begin{eqnarray}
\delta \omega _{\mathrm{qb}} &=&\omega _{\mathrm{qb}}-\omega _{\mathrm{d}},
\\
\delta \omega _{\mathrm{r}} &=&\omega _{\mathrm{r}}-\omega _{\mathrm{d}}.
\notag
\end{eqnarray}

\textbf{Control microwave field.} For the sake of generality, we consider
also the case when the qubit is driven by the separate microwave field,
coupled, for example via an additional microwave line. Then, we have%
\begin{equation}
H_{\mathrm{\mu w}}^{(2)}=-I_{\mathrm{p}}\Phi _{\mathrm{ac}}\cos \omega _{%
\mathrm{d}}t\cdot \sigma _{z},  \label{Hmw2}
\end{equation}%
where $\Phi _{\mathrm{ac}}$ is the amplitude of the driving flux. In the
qubit eigenstate representation, this is simplified to the form of Eq.~(\ref%
{Hmw}) according to
\begin{equation}
H_{\mathrm{\mu w}}^{(2)\prime }=-I_{\mathrm{p}}\Phi _{\mathrm{ac}}\frac{%
e^{i\omega _{\mathrm{d}}t}+e^{-i\omega _{\mathrm{d}}t}}{2}\left( \frac{%
\varepsilon }{\hbar \omega _{\mathrm{qb}}}\sigma _{z}-\frac{\Delta }{\hbar
\omega _{\mathrm{qb}}}\sigma _{x}\right) \approx \xi _{\varepsilon
}(e^{i\omega _{\mathrm{d}}t}\sigma ^{-}+e^{-i\omega _{\mathrm{d}}t}\sigma
^{+}),  \label{Hmw2prime}
\end{equation}%
\begin{equation}
\xi _{\varepsilon }=\frac{1}{2}I_{\mathrm{p}}\Phi _{\mathrm{ac}}\frac{\Delta
}{\hbar \omega _{\mathrm{qb}}}.  \label{ksi_eps}
\end{equation}%
Here we have left only slowly rotating terms (see discussion above). Note
that the amplitude $\xi _{\varepsilon }$ is dependent on the bias $%
\varepsilon $ (see Eq.~(\ref{wqb})). Then in the RWA after the
transformation (\ref{U_RWA}) we obtain the expression which differs from
Eq.~(\ref{H_RWA_2}) by substituting the last term with $\xi _{\varepsilon
}(\sigma ^{+}+\sigma ^{-})$.

\textbf{Dispersive regime. }In the dispersive regime (when $\delta \gg \hbar
g_\epsilon$) the diagonalization of the Hamiltonian (\ref{H_JC}) in the
second order in $\mathrm{g}/\delta $ \cite{Schleich} gives

\begin{equation}
H=-\frac{1}{2}\left( \hbar \omega _{\mathrm{qb}}+\frac{\hbar \mathrm{g}%
_{\varepsilon }^{2}}{\delta }\right) \sigma _{z}+\left( \hbar \omega _{%
\mathrm{r}}+\frac{\hbar \mathrm{g}_{\varepsilon }^{2}}{\delta }\sigma
_{z}\right) a^{\dag }a.  \label{H_JC_diagonlzd}
\end{equation}%
This expression explicitly shows the qubit transition energy shift by the
coupling and also the resonator energy shift by the qubit, which sign
depends on the qubit state.

\section{Solution of the master equation for the density matrix of the system%
}

To describe the qubit-resonator dissipative and incoherent dynamics we
assume that all processes in our system are Markovian and solve the master
equation for the density matrix $\rho $

\begin{equation}
\dot{\rho}=-\frac{i}{\hbar }\left[ H,\rho \right] +\mathcal{L}[\rho ].
\label{1}
\end{equation}%
It includes the dynamic part and dissipative Lindblad term \cite%
{ScullyZubairy}

\begin{equation}
\mathcal{L}[\rho ]=\frac{1}{2}\sum_{k=1}^{3}\left( 2C_{k}\rho C_{k}^{\dag
}-C_{k}^{\dag }C_{k}\rho -\rho C_{k}^{\dag }C_{k}\right) ,  \label{2}
\end{equation}%
where

\begin{align}
C_{1}& =\sqrt{\gamma _{1}}\sigma ,\text{ \ }\gamma _{1}=\frac{1}{T_{1}},
\label{3} \\
C_{2}& =\sqrt{\frac{\gamma _{\phi }}{2}}\sigma _{z},\text{ \ }\gamma _{\phi
}=\frac{1}{T_{\phi }}=\frac{1}{T_{2}}-\frac{1}{2T_{1}},  \notag \\
C_{3}& =\sqrt{\varkappa }a.  \notag
\end{align}%
The Lindblad operator $\mathcal{L}$ presents dissipation in the resonator
(photon decay) with the rate $\varkappa =\varkappa _{\mathrm{ext}}+\varkappa
_{\mathrm{int}}$, where $\varkappa _{\mathrm{ext}}$ and $\varkappa
_{\mathrm{int}}$ are external (leaking out through of the resonator) and internal (resistive) loss rates, and the qubit decoherence consisting of the relaxation
rate $\gamma _{1}$ and the dephasing rate $\gamma _{\phi }$. Here we
consider nondispersive regime (near the qubit-resonator resonance). The
Hamiltonian of the system $H$ in the rotating wave approximation has the
form of Eq.~(\ref{H_RWA_2}). The solution of the master equation determines
the observable quantities, in particular, the expectation value of the
photon field in the resonator
\begin{equation}
\left\langle a\right\rangle =Tr(a\rho ).  \label{5}
\end{equation}

The Hilbert space of the composite system\ is the tensor product of the
qubit space and the photon space with basis vectors $\left\vert
e/g,n\right\rangle =\left\vert e/g\right\rangle \otimes \left\vert
n\right\rangle $. Basis vectors $\left\vert g\right\rangle $\ and \ $%
\left\vert e\right\rangle $

\begin{equation}
\left\vert g\right\rangle =\left[
\begin{array}{c}
0 \\
1%
\end{array}%
\right] ,\text{ }\left\vert e\right\rangle =\left[
\begin{array}{c}
1 \\
0%
\end{array}%
\right]  \label{7}
\end{equation}%
are the eigenvectors of the operator $\sigma _{z}$. Vectors of Fock (photon)
states \ $\left\vert n\right\rangle $\ (the eigenvectors of the photon
number operator $a^{\dag }a\left\vert n\right\rangle =n\left\vert
n\right\rangle $) are the vectors in the infinite-dimensional space $%
N=\infty $\

\begin{equation}
\left\vert 0\right\rangle =\left[
\begin{array}{c}
1 \\
0 \\
0 \\
0 \\
\vdots%
\end{array}%
\right] ,\left\vert 1\right\rangle =\left[
\begin{array}{c}
0 \\
1 \\
0 \\
0 \\
\vdots%
\end{array}%
\right] ,\left\vert 2\right\rangle =\left[
\begin{array}{c}
0 \\
0 \\
1 \\
0 \\
\vdots%
\end{array}%
\right] ,...\left\vert n\right\rangle =\left[
\begin{array}{c}
0 \\
0 \\
\vdots \\
1 \\
\vdots%
\end{array}%
\right] .  \label{8}
\end{equation}%
In the basis $\left\vert e/g,n\right\rangle $ the matrix equation (\ref{1})
is the infinite set of equations for the infinite-dimensional matrix $\rho
_{ij}$.

Below, we consider the simplest case of $N=2$ (weak driving limit), where the analytical solution
is possible, and in the case of $N\gg1$ we study the problem numerically.

\subsection{\textbf{Weak driving limit}}

To find the analytical solution we restrict the photon space to $N=2$,
assuming that the mean photon number in the resonator (created by the
driving field with the amplitude $\xi $) is much less than unity. The basis $%
\left\vert e/g,n\right\rangle $ in this case consists of $4$ base vectors $%
b_{i}$

\begin{equation}
b_{1}=\left\vert g0\right\rangle ,b_{2}=\left\vert e0\right\rangle
,b_{3}=\left\vert g1\right\rangle ,b_{4}=\left\vert e1\right\rangle ,
\label{9}
\end{equation}%
and the density matrix $\rho _{ij}=\left\langle b_{i}\left\vert \rho
\right\vert b_{j}\right\rangle $ takes\ the form

\begin{equation}
\rho =\left(
\begin{array}{cccc}
\rho _{e0,e0} & \rho _{e0,g0} & \rho _{e0,e1} & \rho _{e0,g1} \\
\rho _{g0,e0} & \rho _{g0,g0} & \rho _{g0,e1} & \rho _{g0,e1} \\
\rho _{e1,e0} & \rho _{e1,e0} & \rho _{e1,e1} & \rho _{e1,g1} \\
\rho _{g1,e0} & \rho _{g1,g0} & \rho _{g1,e1} & \rho _{g1,g1}%
\end{array}%
\right) .  \label{10}
\end{equation}

In the steady state from Eq.~(\ref{1}) we have $16$ linear equations for the
matrix elements $\rho _{ij}$. In the weak driving limit, when the drive does
not change population of the ground state $\rho_{g0,g0} = 1$, leaving up to the
first order terms only in the amplitude $\xi $, we obtain the density matrix
$\rho $. The nonzero elements of the matrix $\rho _{ij}$ are

\begin{align}
\rho _{g0,g0}& =1,  \notag \\
\rho _{g1,g0}& =\rho _{g0,g1}^{\ast }=\frac{-i(\xi /\hbar )(\delta \omega _{%
\mathrm{qb}}-i\gamma )}{\mathrm{g}_{\varepsilon }^{2}-(\delta \omega _{%
\mathrm{r}}-i\frac{\varkappa }{2})(\delta \omega _{\mathrm{qb}}-i\gamma )},
\label{11} \\
\rho _{e0,g0}& =\rho _{g0,e0}^{\ast }=\frac{-i(\xi /\hbar )\mathrm{g}%
_{\varepsilon }}{\mathrm{g}_{\varepsilon }^{2}-(\delta \omega _{\mathrm{r}}-i%
\frac{\varkappa }{2})(\delta \omega _{\mathrm{qb}}-i\gamma )},  \notag
\end{align}%
where $\gamma =\frac{\gamma _{1}}{2}+\gamma _{\phi }$.

Using (\ref{10}) in (\ref{5}) we obtain for the mean value of the voltage
field in the resonator in the weak driving (WD) limit for positive
frequencies

\begin{equation}
V_{\mathrm{r0}} \left\langle -i a^\dag\right\rangle _{\mathrm{WD}} = V_{%
\mathrm{r0}}\frac{\xi (\delta \omega _{ \mathrm{qb}}+i\gamma )}{\mathrm{g}%
_{\varepsilon }^{2}-(\delta \omega _{ \mathrm{r}}+i\frac{\varkappa }{2}%
)(\delta \omega _{\mathrm{qb}}+i\gamma )}.  \label{12}
\end{equation}%
The transmission coefficient of the output driving signal $t$\ is defined by
the photon field in the resonator, Eq.~(\ref{tfin}), and according with (\ref%
{12}) we obtain

\begin{equation}
t_{\mathrm{WD}}\mathcal{=}{-i\frac{\varkappa _{\mathrm{ext}}}{2}\frac{\delta
\omega _{\mathrm{qb}}+i\gamma }{\mathrm{g}_{\varepsilon }^{2}-(\delta \omega
_{\mathrm{r}}+i\frac{\varkappa }{2})(\delta \omega _{\mathrm{qb}}+i\gamma )}}%
.  \label{13}
\end{equation}%
When ${\mathrm{g}_{\varepsilon }}=0$, this equation gives the transmission
coefficient through the resonator, which for the linear resonator can be derived classically (Eq.~\ref{trans4}). The plot of the transmission amplitude $\left\vert
t\right\vert _{\mathrm{WD}}$, given by Eq.~(\ref{13}), is shown in Fig.~\ref%
{Fig3} for $\omega _{\mathrm{qb}}=\omega _{\mathrm{r}}$ and different values
of the decay rates $\varkappa $ and $\gamma $ (given in units of the
coupling constant $\mathrm{g}_{\varepsilon }$). For weak decay rates $%
\varkappa $ and $\gamma $, the transmission spectrum displays the
Rabi-splitting peaks (red solid curve), which are smeared with increasing of
the decay.

\begin{figure}[h]
\includegraphics[width=8cm]{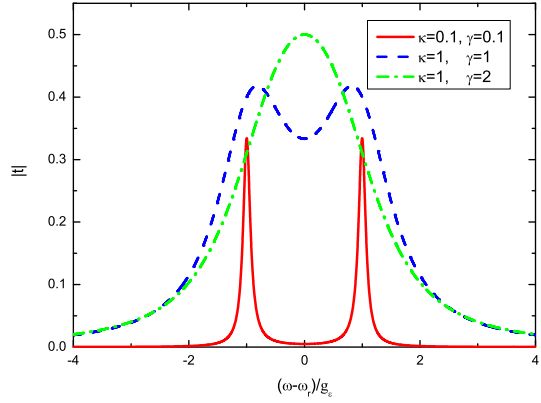}
\caption{Normalized transmission amplitude $\left\vert t\right\vert $ as a
function of the driving frequency detuning $\protect\omega _{\mathrm{d}}-%
\protect\omega _{\mathrm{r}}$ at $\protect\varepsilon =\protect\varepsilon %
^{\ast }$ (when $\protect\omega _{\mathrm{qb}}(\protect\varepsilon ^{\ast })=%
\protect\omega _{\mathrm{r}}$) for different values of the decay rates $%
\varkappa $ and $\protect\gamma $ (given in the figure in units of $\mathrm{g%
}_{\protect\varepsilon }$), calculated with Eq.~(\protect\ref{13}).}
\label{Fig3}
\end{figure}

\begin{figure}[h]
\includegraphics[width=8cm]{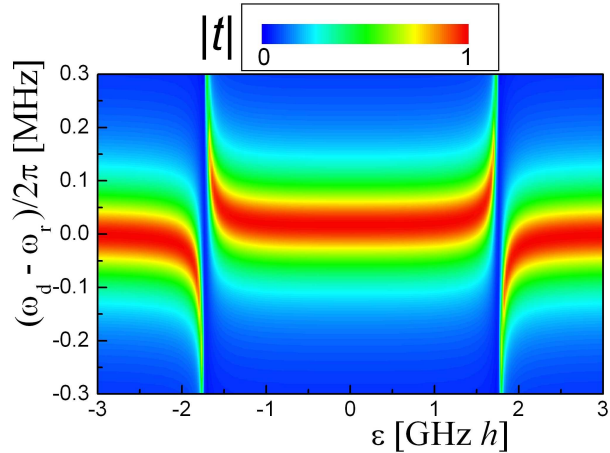}
\caption{Normalized transmission amplitude $\left\vert t\right\vert $ as a
function of the bias $\protect\varepsilon $ and the driving frequency
detuning $\protect\omega _{\mathrm{d}}-\protect\omega _{\mathrm{r}}$,
calculated with Eq.~(\protect\ref{13}).}
\label{Fig4}
\end{figure}
In Fig.~\ref{Fig4} the density plot of the transmission amplitude as a
function of the bias $\varepsilon $ and the detuning $\omega _{\mathrm{d}%
}-\omega _{\mathrm{r}}$ is shown. The parameters here and below are taken
for the comparison with the relevant experimental work \cite{Oelsner09} $%
\Delta /h=1.8$ GHz, $\mathrm{g}/2\pi =3$ MHz, $\omega _{\mathrm{r}}/2\pi
=2.5 $ GHz (the same as in Fig.~\ref{Fig:levels}) and also the loss rate of
the resonator $\varkappa /2\pi =1.25\cdot 10^{-4}$ GHz and the loss rate of
the qubit $\gamma =\mathrm{g}$. Note that we consider the intermediate
coupling regime, when $\mathrm{g}=\gamma \gg \varkappa $. The transmission
amplitude is resonantly increased along the lines shown in Fig.~\ref%
{Fig:levels}(c) as expected. In the narrow vicinity of the resonator
characteristic frequency, $\omega _{\mathrm{d}}\in (\omega _{\mathrm{r}}-%
\mathrm{g},$ $\omega _{\mathrm{r}}+\mathrm{g})$, the avoided crossing at $%
\varepsilon =\varepsilon ^{\ast }$ is demonstrated, as it was reported in
Ref. \cite{Oelsner09}.

\begin{figure}[h]
\includegraphics[width=8cm]{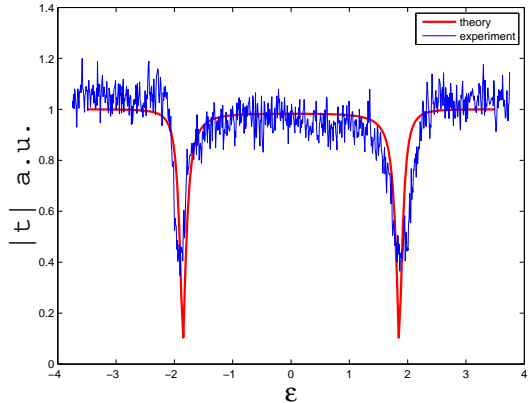}
\caption{Normalized transmission amplitude $\left\vert t\right\vert $ as a
function of the bias $\protect\varepsilon $ for $\protect\omega _{\mathrm{d}%
}=\protect\omega _{\mathrm{r}}$ calculated with Eq.~(\protect\ref{13}), red
line, and obtained experimentally, blue line.}
\label{Fig5}
\end{figure}

For more detailed comparison and finding the parameters with better accuracy
(e.g. decay rate $\gamma $), we need to compare experimental and theoretical
sets of crossection of surfaces $\left\vert t\right\vert $ versus $%
\varepsilon $ and $\omega _{\mathrm{d}}$. This is shown in Fig.~\ref{Fig5}
for $\omega _{\mathrm{d}}=\omega _{\mathrm{r}}$.

\subsection{Numerical solution of the master equation. Beyond the weak
driving regime.}

\begin{figure}[h]
\includegraphics[width=8cm]{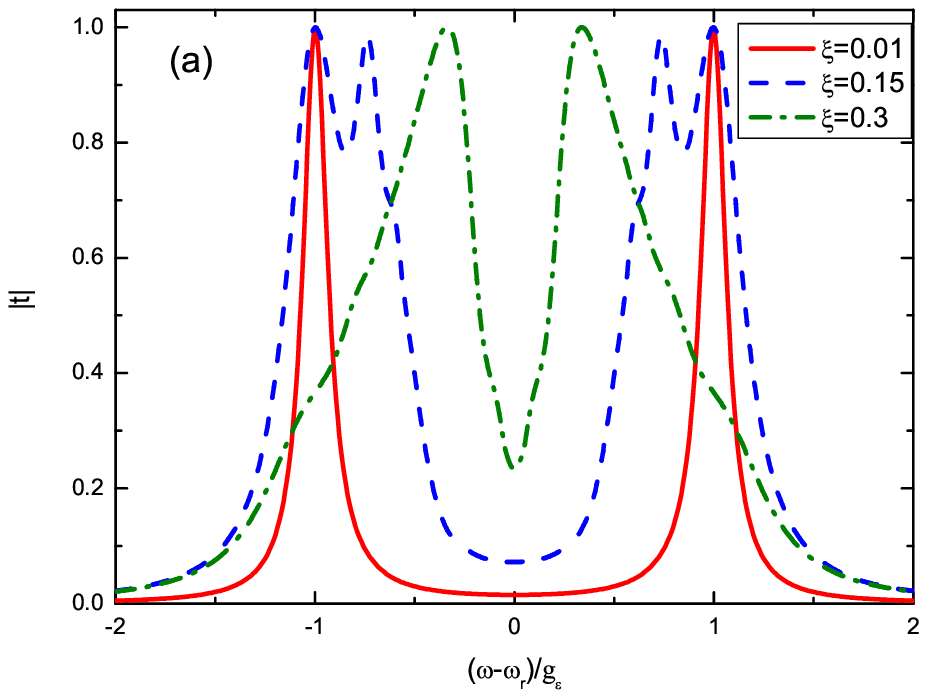} \includegraphics[width=8cm]{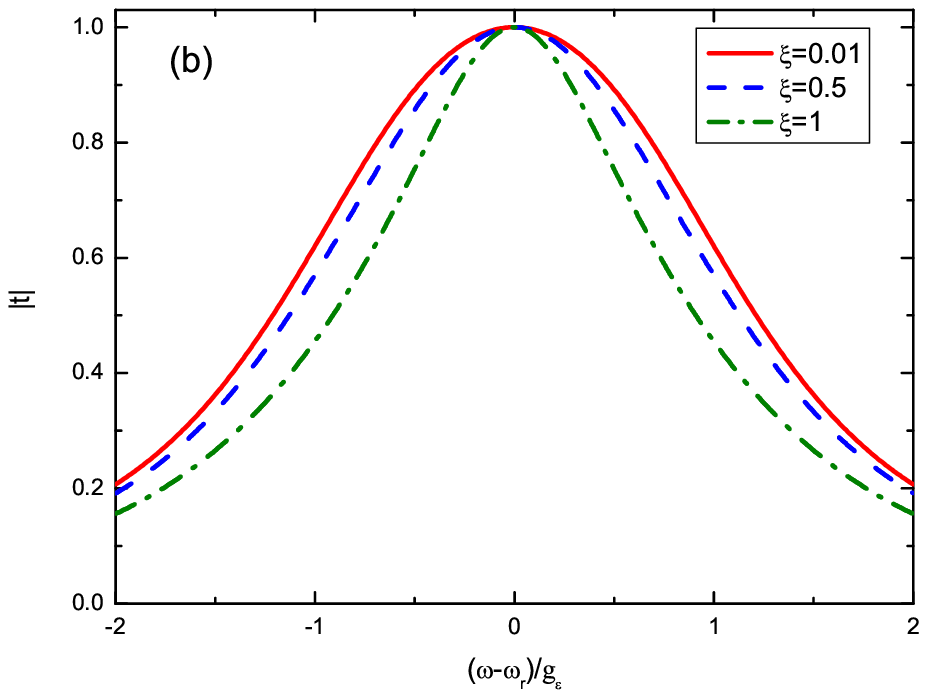}
\caption{Normalized transmission amplitude $\left\vert t\right\vert $ as a
function of the driving frequency detuning $\protect\omega _{\mathrm{d}}-%
\protect\omega _{\mathrm{r}}$ at $\protect\varepsilon =\protect\varepsilon %
^{\ast }$ (when $\protect\omega _{\mathrm{qb}}(\protect\varepsilon ^{\ast })=%
\protect\omega _{\mathrm{r}}$) for (a) $\varkappa /\mathrm{g}_{\protect%
\varepsilon }=0.1$, $\protect\gamma /\mathrm{g}_{\protect\varepsilon }=0.1$
and (b) $\varkappa /\mathrm{g}_{\protect\varepsilon }=1$, $\protect\gamma /%
\mathrm{g}_{\protect\varepsilon }=2$, calculated by solving numerically the
master equation for several values of $\protect\xi $, given in units of $%
\mathrm{g}_{\protect\varepsilon }$.}
\label{Fig6}
\end{figure}

In the case, when driving is not weak, i.e. the mean photon number $%
\left\langle a^{\dag }a\right\rangle \gtrsim 1$, we have solved the equation
for the density matrix $\rho $ numerically. The results are presented in
Fig.~\ \ref{Fig6}. The transmission amplitude $\left\vert t\right\vert $ in
all cases is normalized on the maximal value at $\omega _{\mathrm{qb}%
}=\omega _{\mathrm{r}}$. In Fig.~\ref{Fig6}(a) the transmission amplitude is
shown for the case of small damping $\varkappa /\mathrm{g}_{\varepsilon
}=0.1 $ and $\gamma /\mathrm{g}_{\varepsilon }=0.1$. At a weak driving
amplitude $\xi $ the red curve in Fig.~\ref{Fig6}(a) coincides with $%
\left\vert t\right\vert _{\mathrm{WD}}(\omega _{\mathrm{d}})$ (Fig.~\ref%
{Fig3}). With increasing $\xi $, each split Rabi peak is additionally split
(blue curve) (see also in Ref. \cite{Bishop08}). With further increasing of
the amplitude $\xi $, the additional splitting is smeared (green curve).
Thus in the nonlinear regime we observe the qualitatively new features as
compared to the weak driving limit. When the decay is rather large, such
that in the weak-driving case, we do not have the Rabi splitting (green curve
in Fig.~\ref{Fig3}), in the nonlinear response, we do not observe the
qualitatively new features, as shown in Fig.~\ref{Fig6}(b) ($\varkappa /%
\mathrm{g}_{\varepsilon }=1$ and $\gamma /\mathrm{g}_{\varepsilon }=2$).

We also calculate the average number of photons in the resonator, $%
n=\left\langle a^{\dag }a\right\rangle $. For the parameters in Fig.~\ref%
{Fig6}, it depends on the frequency; the maximal values are the following $%
n_{\max }=0.005$ for $\xi /\mathrm{g}_{\varepsilon }=0.01$, $n_{\max }=0.3$
for $\xi /\mathrm{g}_{\varepsilon }=0.15$, $n_{\max }=1.8$ for $\xi /\mathrm{%
g}_{\varepsilon }=0.3$.

\section{Conclusion}

We presented the detailed theory for the system of the flux qubit coupled
inductively to the transmission line resonator. The transmission coefficient
is calculated with the system's density matrix by solving the master
equation within RWA.

The avoided crossing of the dressed energy levels is shown in the resonant
case, where $\omega _{\mathrm{d}}\approx \omega _{\mathrm{qb}}\approx \omega
_{\mathrm{r}}$. This is demonstrated in the intermediate coupling regime,
which describe the experimental results of Oelsner et al. \cite{Oelsner09}.
We have shown that the dissipation smears the Rabi splitting. Moreover, we
have demonstrated the double splitting in the strong driving regime.

\begin{acknowledgments}
This work was supported by the Fundamental Researches State Fund grant
F28.2/019, by the EU through the EuroSQIP project, by the DFG project IL
150/6-1, by DAAD scholarship A/10/05536. Ya. S. G. and E. I. acknowledge the
financial support from Federal Agency on Science and Innovations of Russian
Federation under contract No. 02.740.11.5067 and the financial support from
Russian Foundation for Basic Research, Grant RFBR-FRSFU No. 09-02-90419. Ya.
S. G. and S. N. Sh. thank P. Macha and G. Oelsner for valuable discussions.
\end{acknowledgments}

%-----------------------------------------------------------
\appendix%-----------------------------------------------------------

\section{Transmission line resonator}

In this Appendix we consider the resonator formed by the transmission line
interrupted by two capacitances $C_{0}$. The qubit we assume to be coupled
inductively to the resonator at its center, see Fig.~\ref{Fig:scheme}(a). We
start by presenting the equations which describe the transmission line.

\subsection{The transmission line}

The transmission line is usually modelled as an infinite series of the
elementary circuits (e.g., \cite{Pozar}), as shown in Fig.~\ref{Fig:scheme}%
(b). Here elementary inductance, capacitance and conductance are $\Delta
L=L\Delta x$, $\Delta C=C\Delta x$, $\Delta G=G\Delta x$, where $L$, $C$ and
$G$ are inductance, capacitance and conductance (of parallel resistance) per
unit length. For the circuit in Fig.~\ref{Fig:scheme}(b), we can
write (neglecting the Ohmic losses) the equations for the transmission line,
by applying the Kirchhoff's laws for the voltage $V(x,t)$ and the current $%
I(x,t)$; in the limit $\Delta x\rightarrow 0$ they take the form%
\begin{eqnarray}
\frac{\partial V(x,t)}{\partial x} &=&-L\frac{\partial I(x,t)}{\partial t},
\label{dVdx} \\
\frac{\partial I(x,t)}{\partial x} &=&-GV(x,t)-C\frac{\partial V(x,t)}{%
\partial t}.
\end{eqnarray}%
These equations can be rewritten for either $I(x,t)$ or $V(x,t)$ as following%
\begin{equation}
\frac{\partial ^{2}A}{\partial x^{2}}-\frac{1}{v^{2}}\frac{\partial ^{2}A}{%
\partial t^{2}}=\frac{\varkappa }{v^{2}}\frac{\partial A}{\partial t},\text{
\ \ }A=\{I,V\},
\end{equation}%
\begin{eqnarray}
v &=&1/\sqrt{LC}, \\
\varkappa &=&G/C.
\end{eqnarray}%
Here $v$ is the phase velocity and $\varkappa $ defines the
loss in the transmission line.

Assuming $I(x,t)=I(x)e^{i\omega t}$ and $V(x,t)=V(x)e^{i\omega t}$, we
obtain
\begin{eqnarray}
\frac{dV(x)}{dx} &=&-i\omega LI(x),  \label{dV/dx} \\
\frac{dI(x)}{dx} &=&-(G+i\omega C)V(x).
\end{eqnarray}%
Then equation for $A(x)=\{I(x),V(x)\}$ can be written as following%
\begin{equation}
\frac{d^{2}A(x)}{dx^{2}}-\gamma ^{2}A(x)=0,
\end{equation}%
\begin{equation}
\gamma =\sqrt{i\omega L(G+i\omega C)}\equiv \alpha +ik.  \label{gamma}
\end{equation}%
Solving equation for $V(x)$ and using Eq.~(\ref{dV/dx}), we obtain%
\begin{equation}
V(x)=V_{0}^{r}e^{-\gamma x}+V_{0}^{l}e^{\gamma x},
\end{equation}%
\begin{equation}
I(x)=\frac{V_{0}^{r}}{Z_{0}}e^{-\gamma x}-\frac{V_{0}^{l}}{Z_{0}}e^{\gamma
x},
\end{equation}%
where
\begin{equation}
Z_{0}=\frac{i\omega L}{\gamma }\equiv Z_{1}+iZ_{2}.
\end{equation}

\begin{equation}
Z_{1}=\frac{\omega Lk}{\alpha ^{2}+k^{2}},\text{ }Z_{2}=\frac{\omega L\alpha
}{\alpha ^{2}+k^{2}}.  \label{Imp}
\end{equation}%
When losses in the line are small ($G\ll \omega C$), we obtain%
\begin{equation}
k\approx \omega \sqrt{LC}=\frac{\omega }{v},\text{ \ }\alpha \approx \frac{G%
}{2}\sqrt{\frac{L}{C}}=\frac{\varkappa }{2v},
\end{equation}%
\begin{equation}
Z_{1}=\sqrt{\frac{L}{C}},\text{ }Z_{2}=\frac{\omega L\alpha }{k^{2}}.
\label{Imp1}
\end{equation}%
Here the constants $V_{0}^{r}$ and $V_{0}^{l}$ are the amplitudes of
the right- and left-moving waves and $Z_{0}$ is the transmission line
characteristic (wave) impedance.

\subsection{Open transmission-line resonator}

We consider the open transmission line of the length $l$. The quality factor
of the resonator \cite{Pozar} can be written as
\begin{equation}
Q=\frac{k}{2\alpha }=\frac{\omega _{\mathrm{r}}C}{G}=\frac{\omega _{\mathrm{r%
}}}{\varkappa }.
\end{equation}%
Now let us define normal modes of the resonator without
dissipation ($\varkappa =0$). Then assuming zero current through the
boundaries at $x = \pm l/2$ for this modes, we obtain%
\begin{equation}
I^{(j)}(x)=\frac{V_{0}^{r}}{Z_{0}}\left(
e^{-ik_{j}x}-(-1)^{j}e^{ik_{j}x}\right) ,
\end{equation}%
\begin{equation}
V^{(j)}(x)=V_{0}^{r}\left( e^{-ik_{j}x}+(-1)^{j}e^{ik_{j}x}\right) ,
\end{equation}%
where $k_{j}l=j\pi ,$ $j=1,2,3,...$. %\begin{eqnarray}
%I_{j}(x) &=&\frac{2V_{0}^{+}}{Z_{0}}\cos k_{j}x, \\
%V_{j}(x) &=&-i2V_{0}^{+}\sin k_{j}x,  \notag \\
%k_{j}l &=&2\pi j-\pi .  \notag
%\end{eqnarray}%
%(The original indexing was confusing, I changed it. In the original version for example V? is the fundamental mode in the resonator (actually V?) and also the wave amplitude at x<-l/2.)
In particular, for the fundamental mode $j=1$ of the resonator we obtain
\begin{equation}
I^{(1)}(x)=\frac{2V_{0}^{r}}{Z_{0}}\cos k_{1}x,
\end{equation}%
\begin{equation}
V^{(1)}(x)=-2iV_{0}^{r}\sin k_{1}x.
\end{equation}%
For the fundamental mode $j=1$ of the $\lambda /2$ resonator ($l=\lambda /2$%
), we have $k_{\mathrm{r}}\equiv k_{1}=\pi /l$, $\omega _{\mathrm{r}}\equiv
\omega _{1}=k_{1}v=\frac{2\pi }{2\sqrt{L_{\mathrm{r}}C_{\mathrm{r}}}}$,
where $L_{\mathrm{r}}=Ll$ and $C_{\mathrm{r}}=Cl$ are the total inductance
and capacitance of the resonator.

Quantization of the resonator eigenmodes results in the following
expressions for the current and voltage operators and the Hamiltonian%
\begin{equation}
\widehat{I} =\sum \sqrt{\frac{\hbar \omega _{j}}{L_{\mathrm{r}}}}\left(
a_{j}+a_{j}^{\dag }\right) \cos k_{j}x,  \label{Current1b}
\end{equation}
\begin{equation}
\widehat{V} =-i\sum \sqrt{\frac{\hbar \omega _{j}}{C_{\mathrm{r}}}}\left(
a_{j}-a_{j}^{\dag }\right) \sin k_{j}x,  \label{Volt1b}
\end{equation}
\begin{equation}
\widehat{H}_{\mathrm{r}}=\sum \hbar \omega _{j}\left( a_{j}^{\dag }a_{j}+%
\frac{1}{2}\right) .
\end{equation}

We consider the frequency close to the fundamental mode frequency $\omega _{%
\mathrm{r}}$, and, therefore, we ignore other modes. For the fundamental
mode, with $k_{1}=\pi /l$ and omitting the index $j=1$, we obtain

\begin{eqnarray}
\widehat{I} &=&I_{\mathrm{r}0}(a+a^{\dag })\cos \frac{\pi x}{l},\text{ \ \ }%
I_{\mathrm{r}0}=\sqrt{\frac{\hbar \omega _{\mathrm{r}}}{L_{\mathrm{r}}}},
\label{I_TL_oper} \\
\widehat{V} &=&iV_{\mathrm{r}0}(a-a^{\dag })\sin \frac{\pi x}{l},\text{ \ \ }%
V_{\mathrm{r}0}=\sqrt{\frac{\hbar \omega _{\mathrm{r}}}{C_{\mathrm{r}}}},
\label{V_TL_oper}
\end{eqnarray}%
where $I_{\mathrm{r}0}$ and $V_{\mathrm{r}0}$ are the zero-point root mean
square (rms) current and voltage, and the Hamiltonian is given by Eq.~(\ref%
{Hosc}).

We also consider the realistic case: the resonator with two point-like
coupling capacitances $C_{0}$ at the ends with $\theta _{1}=\omega
C_{0}Z_{1}\ll 1$.
For the fundamental mode, the current and voltage operators are
modified to
\begin{eqnarray}
\widehat{I} &=&I_{\mathrm{r}0}(a+a^{\dag })\cos k_{\mathrm{r}}x,
\label{I_TL_oper2} \\
\widehat{V} &=&V_{\mathrm{r}0}(ia-ia^{\dag })\sin k_{\mathrm{r}}x.
\label{V_TL_oper2}
\end{eqnarray}%
And we find
\begin{equation}
k_{\mathrm{r}} \approx k_{1}-\frac{2\theta _{1}}{l}
\end{equation}%
from the boundary conditions
\begin{equation}
I_{\mathrm{r}0}(a+a^{\dag })\cos (\pm k_{\mathrm{r}}l/2) =
V_{\mathrm{r}0}[ia (-i\omega C_0)-ia^{\dag }i\omega C_0]\sin (\pm
k_{\mathrm{r}}l/2).
\label{C0_boundary} \\
\end{equation}
This results in the shifted resonant frequency%
\begin{equation}
\omega _{\mathrm{r}}=\omega _{1}\left( 1-\frac{2\theta _{1}}{\pi }\right) ,
\label{wr_shifted}
\end{equation}%
which is slightly lower than the fundamental frequency $\omega _{1}$, due to
external coupling to the outside lines via the
capacitance $C_{0}$.

\subsection{Transmission through the coplanar waveguide resonator}

Now we will consider a classical problem of transmission of waves through the
resonator. It will help us to find correspondence between the classical and
quantum-mechanical solutions and to define the photon decay rates. The
incident wave propagates from left to right and interacts with the
transmission-line resonator at $x=-l/2$ through the capacitance $C_{0}$. The
output wave is detected after another capacitance $C_{0}$ at $x=l/2$. We
will obtain the system of equations for $V_{j}^{r}$ and $V_{j}^{l}$, which
define the classical current and voltage in $j$-th region, $j=1,2,3$,
respectively for $x<-l/2$, $x\in (-l/2,l/2)$, and $x>l/2$;
\begin{equation}
V_{j}(x)=V_{j}^{r}e^{-\gamma (x-x_{j})}+V_{j}^{l}e^{\gamma (x-x_{j})},
\label{Volt1}
\end{equation}%
\begin{equation}
I_{j}(x)=\frac{V_{j}^{r}}{Z_{0}}e^{-\gamma (x-x_{j})}-\frac{V_{j}^{l}}{Z_{0}}%
e^{\gamma (x-x_{j})},  \label{curr1}
\end{equation}%
where $x_{1}=-l/2$, $x_{2}=0$ and $x_{3}=l/2$. We assume the matched
termination (with impedance equal to $Z_{0}$), then there is no left-propagating
wave in the third region, $V_{3}^{l}=0$. The boundary conditions for
currents and voltages at the points $x=\pm l/2$ are the following
\begin{equation}
I_{1}(-l/2)=I_{2}(-l/2),  \label{curr2}
\end{equation}%
\begin{equation}
I_{2}(l/2)=I_{3}(l/2),  \label{curr3}
\end{equation}

\begin{equation}
V_{1}(-l/2)=V_{2}(-l/2)+I_{2}(-l/2)/i\omega C_{0},  \label{Volt2}
\end{equation}%
\begin{equation}
V_{2}(l/2)=V_{3}(l/2)+I_{3}(l/2)/i\omega C_{0}.  \label{Volt3}
\end{equation}%
From these equations, substituting $V_{1}(-l/2)=V_{1}^{r}+V_{1}^{l}$, we
find a useful relation between the field in the resonator and the external
field $V_{3}$
\begin{equation}
V_{3}(l/2)=V_{2}(l/2)\frac{i\theta _{1}}{1+i\theta _{1}},  \label{Volt3b}
\end{equation}%
where $\theta _{1}=\omega C_{0}Z_{1}$.

We define the transmission coefficient $t$ as a ratio between the
transmitted wave and the incident one as
\begin{equation}
t = \frac{V_{3}^{r}}{V_{1}^{r}}  \label{t}
\end{equation}
and find directly from Eqs. (\ref{Volt1}-\ref{Volt2})
\begin{equation}
t=\frac{4\theta_1^{2}e^{-\gamma l}}{4\theta_1^{2}-4i\theta_1 -1+e^{-2\gamma
l}}.  \label{trans}
\end{equation}

For the interesting case of high-Q resonators ($\alpha \ll k$ and $\theta
\approx \omega C_{0}Z_{1}\ll 1$), we can express the transmission
coefficient near the fundamental mode ($\omega _{\mathrm{1}}=v\pi /l$) in
the compact form
\begin{equation}
t\approx \frac{\frac{\varkappa _{\mathrm{ext}}}{\varkappa }}{1-i\frac{%
2\delta \omega }{\varkappa }},  \label{trans4}
\end{equation}%
where $\delta\omega = \omega_{\rm{r}} - \omega_{\rm{d}}$ is detuning from the resonant frequency
\begin{equation}
\omega _{\mathrm{r}}=\omega _{\mathrm{1}}\left( 1-\frac{2\theta _{1}}{\pi }\right)   \label{omegar}
\end{equation}
due to coupling capacitance $C_0$. Note that this formula coincides with Eq. (\ref
{wr_shifted}), however, obtained from the classical solution.
The peak width
\begin{equation}
\Delta \omega =\varkappa ,  \label{width}
\end{equation}%
is determined by the total photon decay rate $\varkappa =\varkappa _{\mathrm{%
ext}}+\varkappa _{\mathrm{int}}$ which is the sum of the photon decay rate
due to the external loss
\begin{equation}
\varkappa _{\mathrm{ext}}=\frac{4\theta _{1}^{2}\omega _{\mathrm{r}}}{\pi }
\label{kappaext}
\end{equation}%
determined by the coupling to the external transmission lines via $C_{0}$
and the internal photon decay rate
\begin{equation}
\varkappa _{\mathrm{int}}=\frac{2\alpha l\omega _{\mathrm{r}}}{\pi }
\label{kappaext}
\end{equation}%
due to dissipations within the resonator. The quality factor is
\begin{equation}
Q=\frac{\omega _{\mathrm{r}}}{\Delta \omega }=\frac{\pi }{4\theta
_{1}^{2}+2\alpha l}.  \label{Q}
\end{equation}%
This rate is consistent with its definition given in Ref. \cite{Oelsner09}.

Below we estimate the photon decay rate $\varkappa $ for the coplanar
waveguide resonator with parameters taken from \cite{Macha} $l=23$ mm, $%
\omega _{\mathrm{r}}/2\pi =2.5$ GHz, $C_{0}=1$fF, $Z_{1}=50$ Ohm, which give
$\theta _{1}=7.8\times 10^{-4}$. The capacitance per unit length $C$ is
calculated from the expression for $\theta _{1}$ at resonance $\theta
_{1}=\pi C_{0}/lC$. We thus obtain $C=1.74\times 10^{-10}$ F/m. Finally, for
the photon decay rate $\varkappa $ we obtain $\varkappa /2\pi =1.95$ kHz.
This value is about two times smaller than the ones obtained in \cite{Macha}%
. We assume that this discrepancy is due to dielectric losses $G$. It allows
us to estimate $\alpha $ from $4\theta _{1}^{2}\approx 2\alpha l$, then $%
\alpha \approx 5.3\times 10^{-5}$ m$^{-1}$. Therefore, for $G$ we obtain $%
G=2\alpha /Z_{1}\approx 2.12$ Ohm$^{-1}$ m$^{-1}$.

\subsubsection{Transmission in the dispersive regime}

Here we consider an effect of the qubit on the transmission coefficient,
substituting the qubit by an additional classical inductance coupled to the
resonator. This classical analogy may be helpful to understand the
quantum-mechanical effect. In the dispersive regime, coupling to the qubit
can be described as an additional classical inductance $L_{\mathrm{qb}}$ at
the position $x=0$. Such a problem is described by adding two more equations
for $x=0$ to the system of equations (\ref{curr2}-\ref{Volt3}), which
follows from Eq.~(\ref{dVdx}) by adding to the r.h.s. the following term
\begin{equation}
-\delta (x)M\frac{\partial I_{\mathrm{qb}}}{\partial t}=-\delta (x)L_{%
\mathrm{qb}}\frac{\partial I(x,t)}{\partial t},
\end{equation}%
where%
\begin{equation}
L_{\mathrm{qb}}=M^{2}\frac{\partial I_{\mathrm{qb}}}{\partial \Phi }.
\end{equation}
In the ground state we have \cite{Greenberg02, Grajcar04}%
\begin{equation}
L_{\mathrm{qb}}=\frac{4M^{2}I_{\mathrm{p}}^{2}\Delta ^{2}}{(\Delta
^{2}+\varepsilon ^{2})^{3/2}}.
\end{equation}

We modify the definition of $V_{2}$ given in Eq. (\ref{Volt1})
%\begin{equation}
\begin{eqnarray}
V_{2l}(x)=V_{2l}^{r}e^{-\gamma x}+V_{2l}^{l}e^{\gamma x} &,&-\frac{l}{2}<x<0,
\label{Volt2m} \\
V_{2r}(x)=V_{2r}^{r}e^{-\gamma x}+V_{2r}^{l}e^{\gamma x} &,&0<x<\frac{l}{2},
\end{eqnarray}%
%
%\end{equation}
then the boundary conditions at $x=0$ are
\begin{eqnarray}
&&I_{2l}(0)=I_{2r}(0), \\
&&V_{2l}(0)=V_{2r}(0)+i\omega L_{\mathrm{qb}}I_{2r}(0).
\end{eqnarray}%
The solution of the system of equations for the transmission coefficient can
be written as
\begin{equation}
t^{\prime }=\left[ \frac{1}{t}-i\frac{\omega L_{\mathrm{qb}}}{8\theta
_{1}^{2}Z_{0}}\left( e^{-\gamma l}-1-i2\theta _{1}\right) ^{2}\right]
^{-1}\approx \left[ \frac{\varkappa }{\varkappa _{\mathrm{ext}}}\left( 1-i%
\frac{2}{\varkappa }\left( \delta \omega +\frac{\omega L_{\mathrm{qb}%
}\varkappa _{\mathrm{ext}}}{4\theta _{1}^{2}Z_{0}}\right) \right) \right]
^{-1}
\end{equation}%
where $t$ is the transmission without the qubit ($L_{\mathrm{q}}=0$) from
Eq. (\ref{trans}). Here we used the following simplifications $-\gamma
l=-ikl-\alpha l=-i\pi -\alpha l$ and $\alpha l\ll 1$, $\theta _{1}\ll 1$.
Finally we rewrite $t^{\prime }$ in the compact form
\begin{equation}
t^{\prime }\approx \frac{\frac{\varkappa _{\mathrm{ext}}}{\varkappa }}{1-i%
\frac{2\delta \omega ^{\prime }}{\varkappa }},  \label{trans4b}
\end{equation}%
where detuning $\delta \omega ^{\prime }=\omega _{\mathrm{r}}^{\prime
}-\omega _{\mathrm{d}}$ from the redefined resonance frequency
\begin{equation}
\omega _{\mathrm{r}}^{\prime }=\omega _{\mathrm{r}}-\frac{\omega _{\mathrm{r}%
}L_{\mathrm{qb}}\varkappa _{\mathrm{ext}}}{4\theta _{1}^{2}Z_{0}}=\omega _{%
\mathrm{r}}\left( 1-\frac{L_{\mathrm{qb}}}{L_{\mathrm{r}}}\right) ,
\end{equation}%
which is shifted due to the extra inductance $L_{\mathrm{qb}}$ in the
resonator. The phase shift of the transmission coefficient $t^{\prime }$ at $%
\delta \omega =0$ is found as
\begin{equation}
\tan \varphi =\frac{\mathrm{Im}[t^{\prime }]}{\mathrm{Re}[t^{\prime }]}=%
\frac{\omega _{\mathrm{r}}L_{\mathrm{qb}}}{2\theta _{1}^{2}Z_{0}}\frac{%
\varkappa _{\mathrm{ext}}}{\varkappa }=\frac{1}{2\pi }\left( \frac{C_{%
\mathrm{r}}}{C_{0}}\right) ^{2}\frac{L_{\mathrm{qb}}}{L_{{\mathrm{r}}}}\frac{%
\varkappa _{\mathrm{ext}}}{\varkappa }.
\end{equation}

In the ground state we obtain%
\begin{eqnarray}
\tan \varphi &=& A\left[ 1+\left( \varepsilon /\Delta \right) ^{2}\right]
^{-3/2}, \\
A &=&\frac{2}{\pi }\left( \frac{C_{\mathrm{r}}}{C_{0}}\right) ^{2}\frac{%
\hbar \mathrm{g}^{2}}{\omega _{\mathrm{r}}\Delta }.
\end{eqnarray}

\subsubsection{Resonant transmission}

The measured resonator field is expressed via the field operator expectation values $%
\langle \widehat{I}\rangle $ or $\langle \widehat{V}\rangle $ (see Eqs. (\ref%
{I_TL_oper}, \ref{V_TL_oper})). Particularly, the expectation value of the
voltage at $x=\pm l/2$ is $V_{\mathrm{r}0}\left\langle ia-ia^{\dag
}\right\rangle \sin {(\pm k_{r}l/2)}\approx \pm V_{\mathrm{r}0}\left\langle
ia-ia^{\dag }\right\rangle $, and the positive frequency component of the
charge on the capacitances $C_{0}$ are
\begin{equation}
\left\langle q^{+}\right\rangle =\pm C_{0}V_{\mathrm{r}0}\left\langle
-ia^{\dag }\right\rangle .  \label{W}
\end{equation}%
The current leaking out from the resonator, expressed via the reflection and transmission
coefficients $r$ and $t$, is a time-derivative of the charge at $x=-l/2$ and $x=l/2$, that is
\begin{equation}
\frac{V_{1}^{r}}{Z_{1}}(1-r) = i\omega \langle q^{+}\rangle .
\end{equation}%
\begin{equation}
\frac{V_{1}^{r}}{Z_{1}}t=i\omega \langle q^{+}\rangle .
\end{equation}%
Then the transmission coefficient can be presented as
\begin{equation}
t = \frac{i\theta V_{\mathrm{r0}}\langle -ia^{\dag }\rangle }{V_{1}^{r}},
\end{equation}%
which after some algebra using Eq. (\ref{xi}) and Eq. (\ref{kappaext}) can
be rewritten in a simple physical form
\begin{equation}
t=-\frac{\varkappa _{\mathrm{ext}}}{2(\xi /\hbar )}\langle a^{\dag
}\rangle .  \label{tfin}
\end{equation}

It is also straightforward to demonstrate that the scattered waves: back scattered $V_1^r r
e^{ik(x+l/2)}$ at $x<-l/2$ and forward $V_1^r t e^{-ik(x-l/2)} - V_1^r e^{-ik(x-l/2)}$ (difference between the transmitted and the undisturbed one as, if there is no resonator) at $x>l/2$ are equal
in amplitude ($1-r = t$) and, therefore, effectively result in zero
interaction energy
\begin{equation}
C_0 V_1^r V_{\mathrm{r0}} \sin{(k_r l/2)} [-r + (1-t)e^{-ik_r l}] = 0,
\label{Hmus2}
\end{equation}
(compare with Eq. (\ref{Hmus})) that is in the quasi-classical approach of scattering, the first order scattering gives an exact solution.

\end{document}